\documentclass{moriond}

\usepackage{amsfonts,amsmath,amssymb}

\def\be{\begin{equation}}
\def\ee{\end{equation}}
\def\bea{\begin{eqnarray}}
\def\eea{\end{eqnarray}}

\newcommand{\Photo}{}

\begin{document}
\vspace*{1cm}
\title{DISFORMAL INVARIANCE OF SECOND ORDER SCALAR-TENSOR THEORIES}

\author{D. BETTONI}

\address{Faculty of Physics, Technion,\\
Haifa, Israel}

\maketitle\abstracts{\textbf{Contribution to the proceedings of the conference "Rencontres de Moriond", Cosmology session, La Thuile, Italy, March 22-29, 2014}\\
${}$\\
The Horndeski action is the most general one involving a metric and a scalar field that leads to second-order field equations in four dimensions. Being the natural extension of the well-known scalar-tensor theories, its structure and properties are worth analysing along the experience accumulated in the latter context. Here, we argue that disformal transformations play, for the Horndeski theory, a similar role to that of conformal transformations for scalar-tensor theories \`a la Brans--Dicke.}

\section{Introduction}

The quest for the understanding of the present day accelerated phase of the universe has produced an outburst of attention towards modified gravity models. Among them a particular role is played by the Horndeski~\cite{Horndeski:1974wa} or Covariant Galileon action~\cite{Deffayet:2009wt}. Thought to be the most general action constructed with a metric and a scalar field able to give second order field equations, it represents a generalization of scalar-tensor theories \`a la Brans--Dicke. Apart from phenomenological implications, more formal questions about this action have proven to be fundamental to a better understanding of its properties: Are there redundancies in its formulation that can be eliminated with a suitable redefinition of the metric? Can we envisage a generalization in the metric that couples to matter? A first attempt to answer these questions is presented below.

\section{The Horndeski action and disformal transformations}

The Horndeski action, rephrased in modern language, is given by
\be
S=\int d^4x\sqrt{-g}\sum_i\mathcal{L}_i+ S_{\text{(m)}}[\bar{g},\psi]\,,
\label{Horndeski}
\ee
\bea
\nonumber
\mathcal{L}_2&=&G_2(\phi,X)\,,\qquad \mathcal{L}_3=G_3(\phi,X)\square\phi\,,
\\
\mathcal{L}_4&=&G_4(\phi,X)R-G_{4,X}(\phi,X)\left[(\square\phi)^2-(\nabla_\mu\nabla_\nu\phi)^2\right]\,,\label{Horndeskicoeff}
\\
\nonumber
\mathcal{L}_5&=&G_5(\phi,X)G_{\mu\nu}\nabla^\mu\nabla^\nu\phi+\frac{G_{5,X}}{6}\left[(\square\phi)^3-3(\square\phi)(\nabla_\nu\nabla_\mu\phi)^2+2(\nabla_\mu\nabla_\nu\phi)^3\right]\,,
\eea
where $X=\nabla_{\mu}\phi\nabla^{\mu}\phi/2$, while matter is coupled to the so called disformal metric~\cite{Bekenstein:1992pj} $\bar{g}_{\mu\nu}=A(\phi,X)g_{\mu\nu}+B(\phi,X)\phi_{,\mu}\phi_{,\nu}$.
This action represents a great generalization not only in its scalar field part but also in the metric matter is coupled to. However, one can still wonder whether suitable metric transformations can be introduced also in this case, leaving the action invariant and linking alternative frame. Indeed, the answer is provided by the particular form of the metric appearing in the matter action.
\subsection*{The general disformal transformation: a dead end or a new starting point?}

The application of the disformal transformation  $\bar{g}_{\mu\nu}=A(\phi,X)g_{\mu\nu}+B(\phi,X)\phi_{,\mu}\phi_{,\nu}$ to the Horndeski action introduces extra terms that cannot be reabsorbed into the Horndeski coefficient functions.~\cite{Bettoni:2013diz,Zumalacarregui:2013pma} One would then expect the second order nature of the equations of motion to be spoiled thus leaving us with the conclusion that such metric transformation is not viable. However, higher-than-second derivatives can be traded for second order derivatives thanks to hidden constraints coming from the field equations.~\cite{Zumalacarregui:2013pma} This fact leaves us with the question if a more general action than Horndeski's that still gives second order field equations may be constructed.

\subsection*{The reduced disformal transformation: framing the Horndeski action}

Given the above results we have investigated the effects on the Horndeski action of the less general disformal transformation $\bar{g}_{\mu\nu}=A(\phi)g_{\mu\nu}+B(\phi)\phi_{,\mu}\phi_{,\nu}$.~\cite{Bettoni:2013diz} Contrarily to the previous case now the Horndeski action is invariant with the effects of the transformation absorbed into redefinitions of the Horndeski coefficients \eqref{Horndeskicoeff} analogously to what happens in scalar-tensor theories. Schematically, we have that the action is mapped as $S[g,\phi;G_i]\rightarrow S[\bar{g},\bar{\phi},\bar{G}_i]$ with $\bar{G}_i(\phi,\bar{X})=f(\phi,\bar X;A,B)G_i(\phi,\bar{X})+g(\phi,\bar{X};G_{j>i},\partial A,\partial B,\partial\partial A,\partial\partial B)$ where the $i=2,3,4,5$. 
This allowed us to prove that an Einstein frame ($G_4=1\,,G_5=0$) can be defined only from an initial Jordan frame action that has $G_5=0$ and $G_4=A(\phi)^2\sqrt{1-2B(\phi)X}$ and that the two frames are equivalent. Moreover, two more intermediate frames, dubbed Galileon and Disformal, can be defined depending if matter is conformally or disformally coupled to the scalar field. This results, being more than just a theoretical exercise, may help in classifying apparently different theories and simplify the investigation of those in different regimes.

\section{Conclusions}
A major tool in physics is represented by symmetries.
This is a clean and precise way to order models, find their
simplest formulations, and identify the minimal set of
degrees of freedom required to fully define a theory. Here, we have presented the effects of disformal transformations on the Horndeski action finding a new
class of scalar-tensor theories of gravity that admits dis-
formally equivalent frames, thus generalizing the
previous results obtained in the context of standard scalar-
tensor theories. This may have important consequences in
cosmological context as it may allow us to identify a large
class of models in different representations of the same
theory, and possibly provide an insight on models beyond Horndeski action.~\cite{Zumalacarregui:2013pma,Gleyzes:2014dya}



\section*{Acknowledgements}

The work presented here has been accomplished during my PhD at SISSA under the supervision of prof. S. Liberati.

\end{document}